\documentclass{article}%
\usepackage{amssymb}
\usepackage{amsmath}
\usepackage{amsfonts}
\usepackage{graphicx}%
\providecommand{\U}[1]{\protect\rule{.1in}{.1in}}
\begin{document}

\title{On the equilibrium state of two rotating charged \\masses in General Relativity }
\author{G.A. Alekseev$^{\text{*)}}$ and V. A. Belinski$^{\text{**)}}$\\$^{\text{*)}}$Steklov Mathematical Institute, Gubkina 8, Moscow 119991,\\Moscow, Russia,\textit{ G.A.Alekseev@mi.ras.ru}\\$^{\text{**)}}$ICRANet, Piazzale della Repubblica, 10, 65122 Pescara, Italy;\\Rome University "La Sapienza", 00185 Rome, Italy;\\IHES, F-91440 Bures-sur-Yvette, France,\textit{ belinski@icra.it }}
\date{}
\maketitle

\begin{abstract}
This paper is a direct continuation of our publication \cite{AB1}

where it was found the exact solution of the Einstein-Maxwell

equations for two static sources of Reissner-Nordstrom type in the \ 

state of the physical equilibrium. Here we present the exact solution 

of these equations for the case of two rotating charged sources 

and we proved the existence of the physical equilibrium state also 

for this general case.

\end{abstract}

\section{Introduction}

In the non-relativistic physics two particles can be in equilibrium if the
product of their masses is equal to the product of their charges (we use the
units $G=c=1$). However, the question on the existence of an analogue of such
equilibrium state in General Relativity is far to be trivial. Besides the
natural mathematical complications, in General Relativity arise two different
types of the "point" centers, namely Black Hole (BH) and Naked Singularity
(NS) and one need to consider all three configurations BH - BH, NS - NS and BH
- NS separately. Yet in each case the notion of a physically sensible distance
between these objects should be defined.

When the Inverse Scattering Method (ISM) have been adopted for integration of
the Einstein and Einstein-Maxwell equations it was shown on the exact
mathematical level that Black Holes and Naked Singularities represent nothing
else but stationary axially symmetric solitons. Then by the ISM machinery one
can obtain the infinite families of exact stationary axially symmetric
solutions of these equations containing such solitons centralized at different
points of the symmetry axis. The formal construction of such solutions do not
represents any difficulties apart of the routine calculations in the framework
of the well developed procedure how to insert a number of solitons into a
given background spacetime. However, it is quite intricate task to single out
from these families the physically reasonable constructions which correspond
to a real equilibrium states of charged Black Holes and Naked Singularities
interacting with each other. The point is that in general the stationary
axially symmetric solitonic solutions possess some features which are
unacceptable from the physical point of view. These unwanted traits are due
the presence in the solutions exotic peculiarities of the following four
types: (i) NUT parameters, (ii) angle deficit at the points of the symmetry
axis, (iii) closed time-like curves around that parts of symmetry axis which
are out of the sources and\ (iv) magnetic charges. The NUT parameters are
incompatible with asymptotic flatness of the spacetime at spatial infinity.
The angle deficit is the well known conical singularity violating the local
Euclidness of space at the points of symmetry axis (it can be treated as some
singular external strut or string preventing the sources to fall onto or to
run away each other). Keeping in mind the physical applications we also should
avoid of any excess of closed timelike curves with respect to those already
existing inside the sources as an inseparable part of their inner structure.
Also magnetic charges should be excluded since their presence contradicts the
Maxwell theory. All four aforementioned phenomena have nothing to do with a
real equilibrium of the physical bodies and the corresponding equilibrium
solution should be free of such pathologies. To deliberate solution from them
one need to place the set of the free parameters of the solution under some
additional restrictions which can be written in the form of some system of
algebraic equations. The problem is that these equations, even for the
simplest case of two objects, are extremely complicated and it is difficult to
resolve them in an exact analytical form in order to see directly whether they
have physically appropriate solutions compatible with the existence of a
positive definite distance between the sources.

However, the aforementioned nuisances constitute the real troubles only in the
general case of the rotating sources. The static case is much more simple and
it would be not an overstatement to say that for the case of 2 non-rotating
charged objects the problem have been solved completely. The first indications
that two static charged masses can stay in real physical equilibrium without
any struts between them and without any other pathologies came from the
results of Bonnor \cite{Bon} and Perry and Cooperstock \cite{Per}. In
\cite{Bon} it was analyzed the equilibrium condition for a charged test
particle in the Reissner-Nordstrom field and it was shown that such test body
can be at rest in the field of the Reissner-Nordstrom source only if they both
are either extreme (the charge equal to mass) or one of them is of BH type
(the charge is less than the mass) and the other is of NS type (the charge is
grater than the mass). There is no way for equilibrium in cases when both
masses are either of NS type or both are of BH type. The more solid arguments
in favour of existence of a static equilibrium configuration for the Black
Hole - Naked Singularity system was presented in \cite{Per}, where both
sources have been treated exactly, that is no one of the components was
considered as test particle. These results have been obtained with the aid of
numerical calculations and three examples of numerical solutions of the
equilibrium equation have been demonstrated. These solutions can correspond to
the equilibrium configurations free of struts, though the authors have not
been able to show the existence of a positive definite distance between the
sources. The authors of \cite{Per} also reported that a number of numerical
experiments for two Black Holes and for two Naked Singularities showed the
negative outcomes, i.e. all tested sets of the parameters was not in power to
satisfy the equilibrium equation. These findings were in agreement with
Bonnor's test particle analysis.

The explicit analytical resolution of the problem in static case have been
presented in paper \cite{AB1} where it was constructed the exact analytic
solution of the Einstein-Maxwell equations for two sources separated by the
well defined positive distance and free of struts or of any other unphysical
properties \footnote{The more details of the derivation of this solution an
interested reader can find in paper \cite{AB2}.}. We showed also that such
solution indeed exists only for the BH-NS\ system and it is impossible to have
the similar equilibrium state for the pair BH-BH or NS-NS. After these results
the natural question arises whether the analogous equilibrium exists for two
rotating sources. It turn out that the answer is affirmative and in the
present paper we demonstrate the exact equilibrium solution of the
Einstein-Maxwell equation for two rotating charged objects one of which is a
Black Hole and another represents a Naked Singularity.

\section{On the general properties of solitonic solutions}

Metric and electromagnetic potentials for any stationary axisymmetric case in
cylindrical Weyl coordinates $\left(  t,\rho,z,\varphi\right)  $ take the
forms:%
\begin{equation}
ds^{2}=-f\left(  d\rho^{2}+dz^{2}\right)  +g_{tt}dt^{2}+2g_{t\varphi
}dtd\varphi+g_{\varphi\varphi}d\varphi^{2}, \label{1}%
\end{equation}%
\begin{equation}
g_{tt}g_{\varphi\varphi}-g_{t\varphi}^{2}=-\rho^{2}. \label{2}%
\end{equation}%
\begin{equation}
A_{t}=A_{t}(\rho,z),\text{ \ }A_{\varphi}=A_{\varphi t}(\rho,z),\text{
\ }A_{\rho}=0,\text{ \ }A_{z}=0\text{ }, \label{3}%
\end{equation}
where all metric coefficients depend only on the variables $\rho,z$ and the
signature of the metric corresponds to the following signs of the metric
coefficients in the diagonal case: $f>0,$ $g_{\varphi\varphi}<0,$ $g_{tt}>0.$
The gravitational solitons (in the context of ISM) as exact solutions of pure
gravity Einstein equations have been introduced in the papers \cite{BZ1, BZ2}.
The generalization of this technique for the coupled gravitational and
electromagnetic fields was constructed in \cite{A1}. Its more detailed
description can be found in \cite{A2} and in the book \cite{BV}. In this
generalized approach one starts from some given background solution of the
Einstein-Maxwell equations and stick into it any desired number of solitons.
We have to do here with the linear spectral differential equations (Lax pair)
for the $3\times3$ matrix function $\Psi(\rho,z,w)$, where $w$ is a complex
spectral parameter independent of coordinates $\rho,z.$ First of all we choose
some background solution of the Einstein-Maxwell equations and find from the
Lax pair the corresponding background spectral matrix $\Psi_{0}(\rho
,z,w).$Using the ISM dressing procedure it is possible to find explicitly the
spectral matrix $\Psi_{n}(\rho,z,w),$ corresponding to the new solution
containing $n$ solitons inserted to the background spacetime and one can
extract the new metric and new electromagnetic potentials from this $\Psi
_{n}.$ The solitonic field added to the background can be characterized by the
matrix $\Psi_{n}\Psi_{0}^{-1}-I$ which is a meromorphic matrix function
tending to zero in the limit $w\rightarrow\infty$ and having $n$ simple poles
in the complex plane of the parameter $w$ (one pole for each soliton).

In pure gravity case some of these poles can be located at the real axis of
the $w$-plane and the corresponding sources have horizons (that is they are of
the BH type) while complex poles generate objects with naked singularities.
However, in the presence of electromagnetic field the formal machinery of the
ISM developed in \cite{A1} in general does not permit for poles to be located
at real axis which means that by this method one can produce solutions
containing sources only of the NS type.\footnote{We said "in general" because
it can be shown that the ISM considered in \cite{A1} can be adjust also to the
real $w$-poles but only for that special restriction on the parameters of the
solutions which correspond to the extreme Black Holes.} Nevertheless, also in
this case after one obtains the final form of solution it is possible to
forget the way how it was derived and to continue the solution analytically in
the space of its parameters in order to get the complete family containing
solutions with real metric of the physical signature and with horizons as
well. However, the technical procedure how to do this is simple only for the
case of one-solitonic solution (that is for the Kerr-Newman case) and some
simple enough generalization of such procedure was found also for two static
solitonic objects \cite{AB3}. In the general case of two rotating sources (the
corresponding 12-parametric solitonic solution have been constructed in
(\cite{A3})) this task is much more complicated. Fortunately, there is an
effective way to get over this difficulty. Because we need to construct
solution of the BH-NS type we can consider the Kerr-Newman black hole as new
background (instead of the flat spacetime) and insert to it one soliton of NS
type. This is exactly what can be done easily with the generating technique
proposed in \cite{A1} and what we are interested in. The exact expressions (in
terms of the Ernst potentials) for the solution together with the proof that
all conditions of the physical equilibrium can be satisfied are given below.

\section{The BH-NS system in a stationary state}

\medskip\noindent Our solution depends on twelve independent real constant
parameters
\begin{equation}
\{m_{{0}},\,a_{{0}},\,b_{0},\,q_{0},\mu_{0}\},\quad\{m_{{s}},\,a_{{s}%
},\,b_{{s}},\,q_{{s}},\mu{}_{{s}}\},\quad l=z_{2}-z_{1}>0\quad\text{and}\quad
c_{0}\text{ }, \label{4}%
\end{equation}
where the parameters with the suffix \textquotedblleft$0$\textquotedblright%
\ are related to the background solution (black hole) and the parameters with
the suffix \textquotedblleft${s}$\textquotedblright\ are the parameters of a
soliton; all parameters are real. The parameter $l$ (which was chosen positive
for definiteness) characterizes a coordinate distance between the sources
because $z_{1}$ and $z_{2}$ determine respectively the location of a black
hole and a naked singularity on the axis. The constant $c_{0}$ is an arbitrary
multiplier in front of the metric coefficient $f$ \ in (\ref{1}) which should
be chosen in accordance, e.g., with the condition of regularity of the axis at
spatial infinity. It is convenient to use two functions of the enumerated
parameters -- a real $\sigma_{0}$ and an pure imaginary $\sigma_{s}$
determined by the relations
\begin{equation}
\sigma_{0}^{2}=m_{0}^{2}+b_{0}^{2}-a_{0}^{2}-q_{0}^{2}-\mu_{0}^{2}\geq
0,\qquad\sigma_{{s}}^{2}=m_{{s}}^{2}+b_{{s}}^{2}-a_{{s}}^{2}-q_{{s}}^{2}%
-\mu_{s}^{2}\leq0. \label{5}%
\end{equation}

Our solution is stationary and axisymmetric and depends on two Weyl
coordinates $\rho,z$. However, it is more convenient to express it in terms of
so called \textquotedblleft bipolar\textquotedblright\ coordinates -- two
pairs of polar coordinate centered respectively at the location of a black
hole (the coordinates with the suffix $1$) and at the location of a naked
singularity (the coordinates with the suffix $2$). Of course, these four
coordinates should satisfy two additional constraints and each of these four
coordinates can be expressed in terms of Weyl coordinates $\rho,z$. The
corresponding defining relations take the forms
\begin{equation}
\rho=\sqrt{x_{1}^{2}-\sigma_{0}^{2}}\sqrt{1-y_{1}^{2}}=\sqrt{x_{2}^{2}%
-\sigma_{{s}}^{2}}\sqrt{1-y_{2}^{2}}\quad\text{and}\quad z=z_{1}+x_{1}%
y_{1}=l+z_{1}+x_{2}y_{2}, \label{6}%
\end{equation}
though, it is worth to note that $z_{1}$ is not an essential parameter because
it determines a shift of the whole configuration of the fields and their
sources along the axis. The inverse relations of bipolar coordinates
$(x_{1},y_{1})$ corresponding to real $\sigma_{0}$ in terms of Weyl
coordinates are
\begin{equation}%
\begin{array}
[c]{l}%
x_{1}=\dfrac{1}{2}\left[  \sqrt{(z-z_{1}+\sigma_{0})^{2}+\rho^{2}}%
+\sqrt{(z-z_{1}-\sigma_{0})^{2}+\rho^{2}}\right]  ,\\[1ex]%
y_{1}=\dfrac{1}{2\sigma_{0}}\left[  \sqrt{(z-z_{1}+\sigma_{0})^{2}+\rho^{2}%
}-\sqrt{(z-z_{1}-\sigma_{0})^{2}+\rho^{2}}\right]  .
\end{array}
\label{7}%
\end{equation}
For the coordinates $(x_{2},y_{2})$ corresponding to imaginary $\sigma_{s}$
($\sigma_{{s}}^{2}<0$) the similar relations are more complicate
($z_{2}=l+z_{1}$):
\begin{equation}%
\begin{array}
[c]{l}%
x_{2}=\sqrt{\dfrac{1}{2}\left[  (z-z_{2})^{2}+\rho^{2}+\sigma_{{s}}%
^{2}\right]  +\sqrt{\dfrac{1}{4}\left[  (z-z_{2})^{2}+\rho^{2}+\sigma_{{s}%
}^{2}\right]  ^{2}-\sigma_{{s}}^{2}(z-z_{2})^{2}}},\\[2ex]%
y_{2}=\dfrac{z-z_{2}}{x_{2}}.
\end{array}
\label{8}%
\end{equation}
Sometimes it is convenient also to use instead of pairs of coordinates
$(x_{1},y_{1})$ and $(x_{2},y_{2})$ the pairs of quasi-spherical coordinates
$(r_{1},\theta_{1})$ and $(r_{2},\theta_{2})$:
\begin{equation}%
\begin{array}
[c]{l}%
x_{1}=r_{1}-m_{0}\text{ },\\
y_{1}=\cos\theta_{1}\text{ }.
\end{array}
\qquad\qquad%
\begin{array}
[c]{l}%
x_{2}=r_{2}-m_{{s}}\text{ },\\
y_{2}=\cos\theta_{2}\text{ }.
\end{array}
\label{9}%
\end{equation}

The Ernst potentials and metric coefficient $f$ of this solution are:%
\begin{equation}
\mathcal{E}=1-\dfrac{2(m_{0}-ib_{0})}{R_{1}}-\dfrac{2(m_{s}-ib_{s})}{R_{2}%
},\qquad\Phi=\dfrac{q_{0}+i\mu_{0}}{R_{1}}+\dfrac{q_{s}+i\mu_{s}}{R_{2}},
\label{10}%
\end{equation}%
\begin{align}
\dfrac{1}{R_{1}}  &  =\dfrac{x_{2}+ia_{s}y_{2}+K_{1}(x_{2}-\sigma_{s}%
y_{2})+L_{1}(x_{1}+\sigma_{0}y_{1})+S_{0}\left(  x_{2}+\sigma_{s}y_{2}\right)
}{D},\label{11}\\
\dfrac{1}{R_{2}}  &  =\dfrac{x_{1}+ia_{0}y_{1}+K_{2}(x_{1}-\sigma_{0}%
y_{1})+L_{2}(x_{2}+\sigma_{s}y_{2})}{D},\nonumber
\end{align}%
\begin{align}
D  &  =(x_{1}+ia_{0}y_{1}+m_{0}-ib_{0})\left[  x_{2}+ia_{s}y_{2}+m_{s}%
-ib_{s}+S_{0}\left(  x_{2}+\sigma_{s}y_{2}\right)  \right]  -\label{12}\\
&  -\left[  m_{0}-ib_{0}-K_{2}(x_{1}-\sigma_{0}y_{1})-L_{2}(x_{2}+\sigma
_{s}y_{2})\right]  \times\nonumber\\
&  \times\left[  m_{s}-ib_{s}-K_{1}(x_{2}-\sigma_{s}y_{2})-L_{1}(x_{1}%
+\sigma_{0}y_{1}\right]  ,\nonumber
\end{align}%
\begin{align}
K_{1}  &  =\dfrac{ia_{s}-\sigma_{s}}{\sigma_{0}+\sigma_{s}+l},\text{ }%
L_{1}=\dfrac{(m_{0}+ib_{0})(m_{s}-ib_{s})-\left(  q_{0}-i\mu_{0}\right)
\left(  q_{s}+i\mu_{s}\right)  }{(ia_{0}-\sigma_{0})(\sigma_{1}+\sigma_{2}%
+l)},\label{13}\\
K_{2}  &  =\dfrac{ia_{0}-\sigma_{0}}{\sigma_{0}+\sigma_{s}-l},\text{ }%
L_{2}=\dfrac{(m_{0}-ib_{0})(m_{s}+ib_{s})-\left(  q_{0}+i\mu_{0}\right)
\left(  q_{s}-i\mu_{s}\right)  }{(ia_{s}-\sigma_{s})(\sigma_{0}+\sigma_{s}%
-l)},\nonumber
\end{align}%
\begin{equation}
S_{0}=-\frac{\sigma_{0}Y_{s}\bar{Y}_{s}}{\sigma_{s}\left(  \sigma_{0}%
^{2}+a_{0}^{2}\right)  (ia_{s}-\sigma_{s})(\sigma_{0}+\sigma_{s}-l)},\text{ }
\label{14}%
\end{equation}%
\begin{equation}
Y_{s}=\left(  m_{0}-ib_{0}\right)  \left(  q_{s}+i\mu_{s}\right)  -\left(
m_{s}-ib_{s}\right)  \left(  q_{0}+i\mu_{0}\right)  , \label{14-1}%
\end{equation}%
\begin{equation}
f=c_{0}\frac{D\bar{D}}{\left(  x_{1}^{2}-\sigma_{0}^{2}y_{1}^{2}\right)
\left(  x_{2}^{2}-\sigma_{s}^{2}y_{2}^{2}\right)  }. \label{15}%
\end{equation}
Here and in what follows the bar over letters means complex conjugation.

\section{Asymptotics at spatial infinity}

It is easy to see that at spatial infinity, i.e. for $\rho^{2}+z^{2}%
\rightarrow\infty,$ the Ernst potentials (10)-(14) have the following
asymptotical behaviour:%
\begin{equation}
\mathcal{E}=1-\dfrac{2(M-iB)}{r}+O(\dfrac{1}{r^{2}}),\qquad\Phi=\dfrac
{Q_{e}+iQ_{m}}{r}+O(\dfrac{1}{r^{2}}),\qquad r=\sqrt{\rho^{2}+z^{2}}\text{ },
\label{16-1}%
\end{equation}
where $M$ and $B$ are the total gravitational mass and total NUT parameter of
the configuration and $Q_{e}$ and $Q_{m}$ are its total electric and magnetic
charges. Simple calculations show that these parameters are:%
\begin{equation}
M=\text{Re}\left[  \dfrac{(m_{0}-ib_{0})(1+K_{1}+L_{1}+S_{0})+(m_{{s}}%
-ib_{{s}})(1+K_{2}+L_{2})}{1+S_{0}-(K_{1}+L_{1})(K_{2}+L_{2})}\right]  ,
\label{16-2}%
\end{equation}%
\begin{equation}
B=-\text{Im}\left[  \dfrac{(m_{0}-ib_{0})(1+K_{1}+L_{1}+S_{0})+(m_{{s}%
}-ib_{{s}})(1+K_{2}+L_{2})}{1+S_{0}-(K_{1}+L_{1})(K_{2}+L_{2})}\right]
\label{17}%
\end{equation}%
\begin{equation}
Q_{e}=\operatorname{Re}\left[  \dfrac{\left(  q_{0}+i\mu_{0}\right)
(1+K_{1}+L_{1}+S_{0})+\left(  q_{s}+i\mu_{s}\right)  (1+K_{2}+L_{2})}%
{1+S_{0}-(K_{1}+L_{1})(K_{2}+L_{2})}\right]  \label{18}%
\end{equation}%
\begin{equation}
Q_{m}=\operatorname{Im}\left[  \dfrac{\left(  q_{0}+i\mu_{0}\right)
(1+K_{1}+L_{1}+S_{0})+\left(  q_{s}+i\mu_{s}\right)  (1+K_{2}+L_{2})}%
{1+S_{0}-(K_{1}+L_{1})(K_{2}+L_{2})}\right]  \label{19}%
\end{equation}
As for the metric coefficient $f$ it should be subject to the constraint
$f\rightarrow1$ in the limit $r\rightarrow\infty.$ This condition can be
satisfied by an appropriate choice of the coefficient $c_{0}$. It is easy to
see that for this we have to choose
\begin{equation}
c_{0}=\left\vert 1+S_{0}-(K_{1}+L_{1})(K_{2}+L_{2})\right\vert ^{-2}\text{ }.
\label{16}%
\end{equation}
Then if we would like to have a configuration without NUT parameter and
magnetic charge we should impose on the parameters of the solution the
restrictions $B=0$ and $Q_{m}=0.$

\section{On the closed time-like curves}

{If at some points of the axis (where }$\rho=0$) one have{\ $g_{t\varphi}%
\neq0$, this implies (in accordance with the relation between the metric
coefficients in Weyl coordinates $g_{tt}g_{\varphi\varphi}-g_{t\varphi}%
^{2}=-\rho^{2}$) that near these points $g_{\varphi\varphi}>0$. Such
inequality means that near these points of the axis the coordinate lines of
the periodic (azimuth angle) coordinate, being closed lines, are time-like. To
avoid such trouble it is necessary to demand that on every part of the axis
$g_{t\varphi}$ should vanish. As it follows directly from the Einstein -
Maxwell equations, on the axis of symmetry $\rho=0$ the value $\Omega
=g_{t\varphi}/g_{tt}$ is independent of $z$ and therefore, it is a constant.
However, this constant can be different on different disconnected parts of the
axis separated by the sources. Therefore, to exclude the existence of closed
time-like curves near the axis, first of all we should impose two conditions
\begin{equation}
\Omega_{-}=\Omega_{i}=\Omega_{+}\text{ }, \label{20}%
\end{equation}
where $\Omega_{-}$, $\Omega_{i}$ and $\Omega_{+}$ are the constants which are
the values of $g_{t\varphi}/g_{tt}$ on the negative, intermediate and positive
parts of the axis respectively. If the conditions (\ref{20}) are satisfied,
the corresponding common value of $\Omega$ can be reduced to zero by a
\textquotedblleft global\textquotedblright\ coordinate transformation of the
form $t^{\prime}=t+a\varphi$, and $\varphi^{\prime}=\varphi$ with an
appropriate constant $a$.{}\footnote{It is worth to note here that this
transformation actually is not an admissible coordinate transformation
(because $\varphi$ is a periodic coordinate but $t$ is not) and its correct
interpretation is connected with some global restruction of the space-time
manifold (some cut-and-past procedure) such that the coordinate lines of
$\varphi$ become non-closed while the other closed lines appear and they
become the coordinate lines of a new asimutal coordinate $\varphi^{\prime}$.}}

In order to satisfy the conditions (\ref{20}) we need to calculate constants
$\Omega_{+}-\Omega_{-}$ and $\Omega_{i}-\Omega_{-}$ and put both of them to
zero. Calculations show that the first constant take simple form:
\begin{equation}
\Omega_{+}-\Omega_{-}=-4B\text{ }, \label{20-0}%
\end{equation}
where $B$ is the total NUT parameter given by the formula (\ref{17}), while
the second parameter is much more complicated:%
\begin{equation}
\Omega_{i}-\Omega_{-}\equiv-4B-\dfrac{\omega_{{\times}}\overline{\omega
}_{{\times}}}{(a_{{\times}}+i\sigma_{{s}})}+\dfrac{(1+2\delta)}{(1-2\delta
)}\dfrac{\mathcal{H}_{0}\overline{\mathcal{H}}_{0}}{(a_{{\times}}+i\sigma
_{{s}})\mathcal{W}_{o}}. \label{20-1}%
\end{equation}
The explicit expression for $\delta$ takes the form:
\begin{equation}%
\begin{array}
[c]{l}%
\delta=\dfrac{\sigma_{0}(m_{0}m_{{s}}+b_{0}b_{{s}}-q_{0}q_{s}-\mu_{0}\mu_{s}%
)}{\sigma_{0}(l^{2}-\sigma_{0}^{2}-\sigma_{{s}}^{2}-2a_{0}a_{{s}}%
)+(l\sigma_{0}+\sigma_{0}^{2}-ia_{0}\sigma_{{s}})(l-\sigma_{0}-\sigma_{{s}%
})S_{0}}\text{ },
\end{array}
\label{27}%
\end{equation}
where $S_{0}$ follows from (\ref{14}) and%
\begin{equation}
\omega_{{\times}}=m_{{\times}}-ib_{{\times}}+i(a_{{\times}}+i\sigma_{{s}%
}),\qquad\mathcal{W}_{o}=(l^{2}-\sigma_{0}^{2}-\sigma_{{s}}^{2})^{2}%
-4\sigma_{0}^{2}\sigma_{{s}}^{2}\text{ }, \label{27-1}%
\end{equation}%
\begin{equation}%
\begin{array}
[c]{l}%
\mathcal{H}_{0}=-2i(l-\sigma_{{s}}-ia_{0}+m_{0}+ib_{0})\overline{X}_{{\times}%
}+\\
(a_{{\times}}+i\sigma_{{s}}-im_{{\times}}+b_{{\times}})[(l-\sigma_{{s}}%
)^{2}-\sigma_{0}^{2}]+\\
+2(a_{{\times}}+i\sigma_{{s}})\left[  (m_{0}+ib_{0})(l-\sigma_{{s}}%
+ia_{0})+\sigma_{0}^{2}+a_{0}^{2}\right]  \text{ }%
\end{array}
\label{20-2}%
\end{equation}%
\begin{equation}
X_{{\times}}=(m_{0}+ib_{0})(m_{{\times}}-ib_{{\times}})-\left(  q_{0}-i\mu
_{0}\right)  \left(  q_{\times}+i\mu_{\times}\right)  . \label{20-2-A}%
\end{equation}
In these formulas we used the new parameters denoted by the same letters but
with subscript "$\times$". These new constants are defined by the relations:%
\begin{equation}
m_{\times}=M-m_{0}\text{ },\text{ }b_{\times}=B-b_{0}\text{ },\text{
}q_{\times}=Q_{e}-q_{0}\text{ },\text{ }\mu_{\times}=Q_{m}-\mu_{0}\text{ },
\label{20-3}%
\end{equation}%
\begin{equation}
a_{\times}+i\sigma_{s}=\frac{\Gamma\bar{\Gamma}\sqrt{c_{0}}}{\left(
a_{s}+i\sigma_{{s}}\right)  \sqrt{\mathcal{W}_{o}}}\text{ }, \label{20-4}%
\end{equation}%
\begin{equation}
\Gamma=\left(  m_{0}+ib_{0}\right)  \left(  m_{{s}}-ib_{{s}}\right)  -\left(
q_{0}-i\mu_{0}\right)  \left(  q_{s}+i\mu_{s}\right)  -\left(  a_{{s}}%
+i\sigma_{{s}}\right)  \left(  a_{0}-i\sigma_{{s}}-il\right)  \text{ },
\label{20-5}%
\end{equation}
where parameters $c_{0,}M,B,Q_{e},Q_{m}$ have been defined earlier by the
relations (\ref{16-2})-(\ref{16}).

\section{On the conical singularity at the axis}

{If conditions (\ref{20}) are satisfied, this does not mean yet that the
geometry on each part of the axis is regular since at the points of different
parts of the axis the local Euclidness of spatial geometry still may occur to
be violated. This behaviour looks like on the surface of a cone near its
vortex, where the ratio of the length of a circle (surrounding the vortex) to
its \textquotedblleft radius\textquotedblright\ is not equal to $2\pi$. In the
solution, on any surface $z=const$ intersecting the axis, the length and
radius of the circles $\rho=const$ are represented asymptotically for
$\rho\rightarrow0$ by the expressions $L=2\pi\sqrt{-g_{\varphi\varphi}}$ and
$R=\sqrt{f}\rho$ respectively. Using the mentioned above relation for metric
coefficients in Weyl coordinates $g_{tt}g_{\varphi\varphi}-g_{t\varphi}%
^{2}=-\rho^{2}$ and the condition $g_{t\varphi}\rightarrow0$ discussed just
above, we obtain that the condition of the local Euclidness of the geometry at
these points, i.e. the condition $L/(2\pi R)\rightarrow1$ for $\rho
\rightarrow0$, is equivalent to the following constraints:
\begin{equation}
P_{-}=P_{i}=P_{+}=1,\qquad P\equiv fg_{tt}\text{ }, \label{21}%
\end{equation}
where $P_{-}$, $P_{i}$ and $P_{+}$ are the values of the product $fg_{tt}$
respectively on the negative, intermediate and positive parts of the axis. (In
accordance with the Einstein - Maxwell equations, the product $fg_{tt}$ is
constant on a part of the axis where }$g_{t\varphi}=0${, however, these
constants again may occur to be different for different disconnected parts of
the axis.) }To obtain conditions which should be imposed on the parameters of
our solution providing the equations (\ref{21}) to be satisfied, we have to
analyze the behaviour of metric components on different parts of the axis of
symmetry. Let's do this.

\noindent$\underline{\text{\textit{Negative semi-infinite part of the axis:}}%
}$ $\{\rho=0,\,-\infty<z<z_{1}-\sigma_{0}\}$. On this part of the axis
\begin{equation}
x_{1}=z_{1}-z,\quad x_{2}=l+z_{1}-z,\quad y_{1}=y_{2}=-1, \label{22}%
\end{equation}
and the direct calculations of the metric component $g_{tt}$ and of the factor
$f$ \ lead to the following expressions%
\begin{align}
g_{tt}  &  =\dfrac{[(z-z_{1})^{2}-\sigma_{0}^{2}][(z-z_{1}-l)^{2}-\sigma_{{s}%
}^{2}]}{c_{0}D_{-}\overline{D}_{-}},\label{23}\\
\text{ }f  &  =\dfrac{c_{0}D_{-}\overline{D}_{-}}{[(z-z_{1})^{2}-\sigma
_{0}^{2}][(z-z_{1}-l)^{2}-\sigma_{{s}}^{2}]}\text{ },\nonumber
\end{align}
where $c_{0}$ has been defined in (\ref{16}) and $D_{-}$ denotes the value of
$D$ (defined by (\ref{12})) on the negative semi-infinite part of the axis.
From these expressions we see that the condition (\ref{21}) is satisfied
automatically on the negative semi-infinite part of the axis. \bigskip

\noindent$\underline{\text{\textit{Positive semi-infinite part of the axis:}}%
}$ \quad$\{\rho=0,\,z_{1}+l<z<\infty\}$. On this part of the axis
\begin{equation}
x_{1}=z-z_{1},\quad x_{2}=z-z_{1}-l,\quad y_{1}=y_{2}=1, \label{24}%
\end{equation}
and here for the metric component $g_{tt}$ and the conformal factor we have
the similar expressions%
\begin{align}
g_{tt}  &  =\dfrac{[(z-z_{1})^{2}-\sigma_{0}^{2}][(z-z_{1}-l)^{2}-\sigma_{{s}%
}^{2}]}{c_{0}D_{+}\overline{D}_{+}},\label{24-1}\\
f  &  =\dfrac{c_{0}D_{+}\overline{D}_{+}}{[(z-z_{1})^{2}-\sigma_{0}%
^{2}][(z-z_{1}-l)^{2}-\sigma_{{s}}^{2}]}\text{ },\nonumber
\end{align}
where $D_{+}$ denotes the value of $D$ on the positive semi-infinite part of
the axis. From these expressions we see that the condition (\ref{21}) is
satisfied identically also on the positive semi-infinite part of the axis.

\noindent$\underline{\text{\textit{Intermediate part of the axis:}}}$
\quad$\{\rho=0,\,z_{1}+\sigma_{0}<z<z_{1}+l\}$. On this part of the axis
\begin{equation}
x_{1}=z-z_{1},\quad x_{2}=l+z_{1}-z,\quad y_{1}=1,\quad y_{2}=-1, \label{25}%
\end{equation}
and the corresponding expressions for $g_{tt}$ and the coefficient $f$ on the
axis take the forms:
\begin{align}
g_{tt}  &  =\dfrac{[(z-z_{1})^{2}-\sigma_{0}^{2}][(z-z_{1}-l)^{2}-\sigma_{{s}%
}^{2}]}{c_{0}D_{i}\overline{D}_{i}}\left(  \dfrac{1-2\delta}{1+2\delta
}\right)  ^{2},\text{ \ }\label{26}\\
f  &  =\dfrac{c_{0}D_{i}\overline{D}_{i}}{[(z-z_{1})^{2}-\sigma_{0}%
^{2}][(z-z_{1}-l)^{2}-\sigma_{{s}}^{2}]}\text{ },\nonumber
\end{align}
where parameter function $\delta$ is the same which have been defined already
by the formula (\ref{27}). As follows from these last expression the condition
(\ref{21}) on the intermediate part of the axis (i.e. the equation $P=1$) is
equivalent to the constraint $\delta=0$.

\section{Physical magnetic and electric charges}

To obtain a physical result we have to exclude from the solution magnetic
charges of both sources. To calculate the physical values of magnetic charges
(note that formal parameters $\mu_{0}$ and $\mu_{s}$ are not physical magnetic
charges, they coincide with them only in the limit of infinite distance
between the sources $l\rightarrow\infty$) we should consider the magnetic
fluxes coming from closed space-like surfaces surrounding each charged center
and apply the Gauss theorem. In this way we can find the physical magnetic
charges $\mu_{0}^{(ph)},\mu_{s}^{(ph)}$ (as well as physical electric charges
$q_{0}^{(ph)},q_{s}^{(ph)}$ ) of each source calculating the corresponding
Komar-like integrals. The detailed procedure how to do this have been
described in the section "Physical parameter of the sources" in paper
\cite{AB2}. The results of these calculations are:%
\begin{equation}
q_{0}^{(ph)}=q_{0}+\operatorname{Re}F\text{ },\text{ }\mu_{0}^{(ph)}=\mu
_{0}+\operatorname{Im}F\text{ }, \label{28}%
\end{equation}%
\begin{equation}
q_{s}^{(ph)}=q_{\times}-\operatorname{Re}F\text{ },\text{ }\mu_{s}^{(ph)}%
=\mu_{\times}-\operatorname{Im}F\text{ }, \label{29}%
\end{equation}
where
\begin{equation}
F=\left(  q_{\times}+i\mu_{\times}\right)  \dfrac{(a_{{\times}}+i\sigma_{{s}%
}-im_{{\times}}+b_{{\times}})}{2(a_{{\times}}+i\sigma_{{s}})}-\dfrac
{\mathcal{L}_{0}\mathcal{H}_{0}}{2\mathcal{W}_{o}(a_{{\times}}+i\sigma_{{s}}%
)}\dfrac{(1+2\delta)}{(1-2\delta)}\text{ }, \label{30}%
\end{equation}
and we introduced here the new parameter polynomial:%

\begin{gather}
\mathcal{L}_{0}=\left[  (l+\sigma_{{s}})^{2}-\sigma_{0}^{2}\right]  \left(
q_{\times}+i\mu_{\times}\right)  +\label{31}\\
+2\left[  X_{{\times}}-i(a_{{\times}}+i\sigma_{{s}})(l+\sigma_{{s}}%
-ia_{0})\right]  \left(  q_{0}+i\mu_{0}\right)  .\nonumber
\end{gather}
The physically acceptable solution corresponds to the restrictions $\mu
_{0}^{(ph)}=\mu_{s}^{(ph)}=0.$

\section{Summary for the equilibrium conditions}

Here we present a summary of the conditions which provide an equilibrium of
two interacting Kerr-Newman sources and which should be satisfied by the
parameters of the whole configuration. Thus, if we fix an arbitrary constant
multiplier in the metric coefficient $f$ as it was described above, we have an
eleven-parameter solution. The first of the equilibrium conditions is the
vanishing of a NUT parameter $B$. Then from (\ref{17}) we have:
\begin{equation}
\text{Im}\left[  \dfrac{(m_{0}-ib_{0})(1+K_{1}+L_{1}+S_{0})+(m_{{s}}-ib_{{s}%
})(1+K_{2}+L_{2})}{1+S_{0}-(K_{1}+L_{1})(K_{2}+L_{2})}\right]  =0 \label{Eq1}%
\end{equation}
The second condition provides the local Euclidness of the geometry on the axis
(i.e. the absence of the conical points) that is equivalent to the vanishing
of the parameter $\delta$ which was defined by the expression (\ref{27}):
\begin{equation}
m_{0}m_{{s}}+b_{0}b_{{s}}=q_{0}q_{s}+\mu_{0}\mu_{s} \label{Eq2}%
\end{equation}
The third condition follows from the absence of the closed time-like curves.
This demand consists of two restrictions $\Omega_{+}-\Omega_{-}=0$ and
$\Omega_{i}-\Omega_{-}$ $=0.$ Taking into account equations (\ref{Eq1}) and
(\ref{Eq2}), these two restrictions gives now only one independent relation
$\Omega_{i}-\Omega_{-}$ $=0$ which acquire the form (see (\ref{20-0}) and
(\ref{20-1})):
\begin{equation}
\mathcal{W}_{o}\omega_{{\times}}\overline{\omega}_{{\times}}=\mathcal{H}%
_{o}\overline{\mathcal{H}}_{o} \label{Eq3}%
\end{equation}
And finally we have to eliminate the physical magnetic charges $\mu_{0}%
^{(ph)},\mu_{s}^{(ph)}$of the sources. In accordance with (\ref{28}%
)-(\ref{31}), bearing in mind that $\delta=0$ in agreement with (\ref{Eq2}),
we have to add the following two conditions:

\smallskip%
\begin{equation}
\mu_{0}+\operatorname{Im}\left[  \left(  q_{\times}+i\mu_{\times}\right)
\dfrac{(a_{{\times}}+i\sigma_{{s}}-im_{{\times}}+b_{{\times}})}{2(a_{{\times}%
}+i\sigma_{{s}})}-\dfrac{\mathcal{L}_{0}\mathcal{H}_{0}}{2\mathcal{W}%
_{o}(a_{{\times}}+i\sigma_{{s}})}\right]  \text{ }=0 \label{Eq4}%
\end{equation}%
\begin{equation}
\mu_{\times}-\operatorname{Im}\left[  \left(  q_{\times}+i\mu_{\times}\right)
\dfrac{(a_{{\times}}+i\sigma_{{s}}-im_{{\times}}+b_{{\times}})}{2(a_{{\times}%
}+i\sigma_{{s}})}-\dfrac{\mathcal{L}_{0}\mathcal{H}_{0}}{2\mathcal{W}%
_{o}(a_{{\times}}+i\sigma_{{s}})}\right]  =0 \label{Eq5}%
\end{equation}

\section{Evidence of the existing of the physical equilibrium}

The easiest way to prove that equilibrium equations have physically acceptable
solutions is to consider the limit of large distance between the sources.
Assuming that $l$ is much larger than all other parameters we can try to find
an expansion for the solution of equations (\ref{Eq1})-(\ref{Eq5}) with
respect to the quantity $1/l.$ It turns out that solution in the form of such
expansion indeed exists. To show this we assume that six parameters $m_{{0}%
},a_{{0}},\,q_{0},m_{{s}},a_{{s}},\,q_{{s}}$ are independent of $l$ while the
four constants $\,b_{0},\mu_{0},b_{{s}},\mu{}_{{s}}$ follows from the
equilibrium equations (\ref{Eq1})-(\ref{Eq5}) as functions of $l$ and of the
aforementioned six parameters $m_{{0}},a_{{0}},\,q_{0},m_{{s}},a_{{s}%
},\,q_{{s}}.$ Then, in the limit of large $l,$ these functions can be expanded
with respect to $1/l.$ The calculations show that these expansions are:%
\begin{equation}
b_{0}=\frac{m_{0}\left(  a_{s}+i\sqrt{m_{s}^{2}-a_{s}^{2}-q_{s}^{2}}\right)
}{l}+O\left(  \dfrac{1}{l^{2}}\right)  , \label{32}%
\end{equation}%
\begin{equation}
\mu_{0}=-\frac{q_{0}\left(  a_{s}+i\sqrt{m_{s}^{2}-a_{s}^{2}-q_{s}^{2}%
}\right)  }{l}+O\left(  \dfrac{1}{l^{2}}\right)  , \label{33}%
\end{equation}%
\begin{equation}
b_{s}=-\frac{m_{s}a_{0}}{l}+O\left(  \dfrac{1}{l^{2}}\right)  , \label{34}%
\end{equation}%
\begin{equation}
\mu_{s}=\frac{q_{s}a_{0}}{l}+O\left(  \dfrac{1}{l^{2}}\right)  . \label{35}%
\end{equation}
Here, in accordance with the initial inequalities (\ref{5}), $m_{s}^{2}%
-a_{s}^{2}-q_{s}^{2}<0$ then all parameters (\ref{32})-(\ref{35}) are real as
it should be. Using this result it is easy to obtain from (\ref{16-2}) and
(\ref{18}) the total mass and total electric charge of the configuration:
\begin{equation}
M=m_{0}+m_{s}+\frac{m_{s}\left(  \sqrt{m_{0}^{2}-a_{0}^{2}-q_{0}^{2}}%
-\Lambda\right)  }{l}+O\left(  \dfrac{1}{l^{2}}\right)  , \label{36}%
\end{equation}%
\begin{equation}
Q_{e}=q_{0}+q_{s}+\frac{q_{s}\left(  \sqrt{m_{0}^{2}-a_{0}^{2}-q_{0}^{2}%
}-\Lambda\right)  }{l}+O\left(  \dfrac{1}{l^{2}}\right)  . \label{37}%
\end{equation}
where
\begin{equation}
\Lambda=-\frac{\sqrt{m_{0}^{2}-a_{0}^{2}-q_{0}^{2}}\left(  m_{0}q_{s}%
-m_{s}q_{0}\right)  ^{2}}{\left(  m_{0}^{2}-q_{0}^{2}\right)  \left(
m_{s}^{2}-a_{s}^{2}-q_{s}^{2}-i\sqrt{m_{s}^{2}-a_{s}^{2}-q_{s}^{2}}\right)
}\text{ }. \label{38}%
\end{equation}
We remind that, again in accordance with the inequalities (\ref{5}), the
quantity $m_{0}^{2}-a_{0}^{2}-q_{0}^{2}$ is positive.

There is one additional important result which is coming from the solution of
the system (\ref{Eq1})-(\ref{Eq5}): the basic (independent of $l$) parameters
$m,\,q_{0},m_{{s}},\,q_{{s}}$ must satisfy the relation%
\begin{equation}
m_{0}m_{{s}}=q_{0}q_{s}\text{ }. \label{39}%
\end{equation}
This equation follows from (\ref{Eq2}), that is \ from the absence of the
conic syngularities at the points of symmetry axis. Then the total amount of
the equilibrium constraints on the initial 11 parameters $m_{{0}},\,a_{{0}%
},\,b_{0},\,q_{0},\mu_{0},m_{{s}},\,a_{{s}},\,b_{{s}},\,q_{{s}},\mu{}_{{s}},l$
are (\ref{32})-(\ref{35}) and (\ref{39}), that is five conditions from five
equations (\ref{Eq1})-(\ref{Eq5}) as it should be. Consequently the solution
describing the physical equilibrium state of two charged rotating sources
contains six arbitrary constants. This is by two constants more then the
corresponding static solution we found in \cite{AB1} which is natural since in
case of rotating sources the two rotation parameters $a_{0}$ and $a_{s}$ appeared.

It is worth to mention that in derivation of the expressions (\ref{32}%
)-(\ref{38}) we used the condition (\ref{39}) which lead to the essential
simplifications of these formulas. Another remark we would like to add is the
fact that formulas (\ref{32})-(\ref{37}) we obtained up to the order
$1/l^{3}.$ These additional details showed no surprises and there is no of big
interest to exhibit them here.

The last remark relates to the total angular momentum of the configuration. To
find it it is necessary to expand the Ernst potential $\mathcal{E}$ at spatial
infinity up to the order $1/r^{2}$ choosing an appropriate spherical
coordinates $r,\theta.$ Then the angular momentum $J$ reveal itself in the
summand $2iJr^{-2}\cos\theta$ in $\mathcal{E}$. We calculated it but the
expression is rather long and we will not show it here. The only interesting
fact is that in the limit of large $l,$ discussed above, the total angular
momentum is%
\begin{equation}
J=m_{0}a_{0}+m_{s}a_{s}+O\left(  \dfrac{1}{l}\right)  . \label{40}%
\end{equation}
This result together with relations $M=m_{0}+m_{s}$ and $Q_{e}=q_{0}+q_{s}$
following from (\ref{36})-(\ref{37}) in the first non-vanishing approximation
and relation $m_{0}m_{{s}}=q_{0}q_{s}$ show that at the infinite separation
between the sources they are just usual electrically charged rotating
Kerr-Newman objects without any NUT annoyance and without magnetic charges.

Of course the analysis presented in this section can be consider as a proof of
existence of the equilibrium for the distances $l$ from the region
$l_{cr}<l<\infty$ but the critical value $l_{cr}$ can be extracted only from
the exact solution of the equilibrium equations (\ref{Eq1})-(\ref{Eq5}) (it is
probable that in general $l_{cr}$ is simply zero under an appropriate
definition of the distance). This task we postpone for a future work.

\section{Acknowledgements}

GAA express his deep thanks to ICRANet (Pescara, Italy) and to the Institut
des Hautes Etudes Scientifiques (Bures-sur-Yvette, France) for the financial
support and hospitality during his visits in June and in October 2012
respectively, when the parts of this work were made. The work of GAA was
supported in parts by the Russian Foundation for Basic Research (grants
11-01-00034, 11-01-00440) and the program "Fundamental problems of Nonlinear
Dynamics" of Russian Academy of Sciences.

\end{document}